\input harvmac\skip0=\baselineskip

\def\p{\partial}

\def\msurr{\mathsurround=0pt}
\def\overleftrightarrow#1{\vbox{\msurr\ialign{##\crcr
        $\leftrightarrow$\crcr\noalign{\kern-1pt\nointerlineskip}
        $\hfil\displaystyle{#1}\hfil$\crcr}}}

\def\frac#1#2{{#1\over #2}}

\def\apm{{\alpha^{\prime}}}

\def\lra#1{{\langle #1\rangle}}

\def\ket#1{{| #1 \rangle}}

\def\ads{{AdS_3}}
\def\SL{{SL(2,R)}}
\def\SU{{SU(2)}}
\def\CSL{{\widehat{SL}(2,R)}}
\def\CSU{{\widehat{SU}(2)}}

\lref\AharonyTI{
O.~Aharony, S.~S.~Gubser, J.~M.~Maldacena, H.~Ooguri and Y.~Oz,
``Large N field theories, string theory and gravity,''
Phys.\ Rept.\  {\bf 323}, 183 (2000)
[arXiv:hep-th/9905111].
}

\lref\MaldacenaBW{
J.~M.~Maldacena and A.~Strominger,
``AdS(3) black holes and a stringy exclusion principle,''
JHEP {\bf 9812}, 005 (1998)
[arXiv:hep-th/9804085].
}

\lref\MaldacenaRE{
J.~M.~Maldacena,
``The large N limit of superconformal field theories and supergravity,''
Adv.\ Theor.\ Math.\ Phys.\  {\bf 2}, 231 (1998)
[Int.\ J.\ Theor.\ Phys.\  {\bf 38}, 1113 (1999)]
[arXiv:hep-th/9711200].
}

\lref\WittenQJ{
E.~Witten,
``Anti-de Sitter space and holography,''
Adv.\ Theor.\ Math.\ Phys.\  {\bf 2}, 253 (1998)
[arXiv:hep-th/9802150].
}

\lref\GubserBC{
S.~S.~Gubser, I.~R.~Klebanov and A.~M.~Polyakov,
``Gauge theory correlators from non-critical string theory,''
Phys.\ Lett.\ B {\bf 428}, 105 (1998)
[arXiv:hep-th/9802109].
}

\lref\thooft{
G.~'t Hooft, 
``A Planar Diagram Theory For Strong Interactions,''
Nucl.\ Phys.\ B {\bf 72}, 461 (1974)
}

\lref\MetsaevBJ{
R.~R.~Metsaev,
``Type IIB Green-Schwarz superstring in plane wave Ramond-Ramond  background,''
Nucl.\ Phys.\ B {\bf 625}, 70 (2002)
[arXiv:hep-th/0112044].
}

\lref\moo{
J.~M.~Maldacena and H.~Ooguri,
``Strings in AdS(3) and SL(2,R) WZW model. I: the Spectrum,''
J.\ Math.\ Phys.\  {\bf 42}, 2929 (2001)
[arXiv:hep-th/0001053].
}

\lref\mos{
J.~M.~Maldacena, H.~Ooguri and J.~Son,
``Strings in AdS(3) and the SL(2,R) WZW model. II: Euclidean black hole,''
J.\ Math.\ Phys.\  {\bf 42}, 2961 (2001)
[arXiv:hep-th/0005183].
}

\lref\mot{
J.~M.~Maldacena and H.~Ooguri,
``Strings in AdS(3) and the SL(2,R) WZW model. III: Correlation  functions,''
Phys.\ Rev.\ D {\bf 65}, 106006 (2002)
[arXiv:hep-th/0111180].
}

\lref\ArgurioTB{
R.~Argurio, A.~Giveon and A.~Shomer,
``Superstrings on AdS(3) and symmetric products,''
JHEP {\bf 0012}, 003 (2000)
[arXiv:hep-th/0009242].
}

\lref\bmn{
D.~Berenstein, J.~M.~Maldacena and H.~Nastase,
``Strings in flat space and pp waves from N = 4 super Yang Mills,''
JHEP {\bf 0204}, 013 (2002)
[arXiv:hep-th/0202021].
}

\lref\BerensteinSA{
D.~Berenstein and H.~Nastase,
``On lightcone string field theory from super Yang-Mills and holography,''
[arXiv:hep-th/0205048].
}

\lref\KiemXN{
Y.~j.~Kiem, Y.~b.~Kim, S.~m.~Lee and J.~m.~Park,
``pp-wave / Yang-Mills correspondence: An explicit check,''
Nucl.\ Phys.\ B {\bf 642}, 389 (2002)
[arXiv:hep-th/0205279].
}

\lref\LeeRM{
P.~Lee, S.~Moriyama and J.~w.~Park,
``Cubic interactions in pp-wave light cone string field theory,''
Phys.\ Rev.\ D {\bf 66}, 085021 (2002)
[arXiv:hep-th/0206065].
}

\lref\ConstableVQ{
N.~R.~Constable, D.~Z.~Freedman, M.~Headrick and S.~Minwalla,
``Operator mixing and the BMN correspondence,''
JHEP {\bf 0210}, 068 (2002)
[arXiv:hep-th/0209002].
}

\lref\BeisertBB{
N.~Beisert, C.~Kristjansen, J.~Plefka, G.~W.~Semenoff and M.~Staudacher,
``BMN correlators and operator mixing in N = 4 super Yang-Mills theory,''
Nucl.\ Phys.\ B {\bf 650}, 125 (2003)
[arXiv:hep-th/0208178].
}
\lref\LeeVZ{
P.~Lee, S.~Moriyama and J.~w.~Park,
``A note on cubic interactions in pp-wave light cone string field  theory,''
[arXiv:hep-th/0209011].
}

\lref\GepnerWI{
D.~Gepner and E.~Witten,
``String Theory On Group Manifolds,''
Nucl.\ Phys.\ B {\bf 278}, 493 (1986).
}

\lref\BalogJB{
J.~Balog, L.~O'Raifeartaigh, P.~Forgacs and A.~Wipf,
``Consistency Of String Propagation On Curved Space-Times: An SU(1,1) Based Counterexample,''
Nucl.\ Phys.\ B {\bf 325}, 225 (1989).
}

\lref\EvansQU{
J.~M.~Evans, M.~R.~Gaberdiel and M.~J.~Perry,
``The no-ghost theorem for AdS(3) and the stringy exclusion principle,''
Nucl.\ Phys.\ B {\bf 535}, 152 (1998)
[arXiv:hep-th/9806024].
}

\lref\IsraelRY{
D.~Israel, C.~Kounnas and M.~P.~Petropoulos,
``Superstrings on NS5 backgrounds, deformed AdS(3) and holography,''
JHEP {\bf 0310}, 028 (2003)
[arXiv:hep-th/0306053].
}

\lref\MaldacenaUZ{
J.~M.~Maldacena, J.~Michelson and A.~Strominger,
``Anti-de Sitter fragmentation,''
JHEP {\bf 9902}, 011 (1999)
[arXiv:hep-th/9812073].
}

\lref\SeibergXZ{
N.~Seiberg and E.~Witten,
``The D1/D5 system and singular CFT,''
JHEP {\bf 9904}, 017 (1999)
[arXiv:hep-th/9903224].
}

\lref\MaldacenaUZ{
J.~M.~Maldacena, J.~Michelson and A.~Strominger,
``Anti-de Sitter fragmentation,''
JHEP {\bf 9902}, 011 (1999)
[arXiv:hep-th/9812073].
}

\lref\SpradlinAR{
M.~Spradlin and A.~Volovich,
``Superstring interactions in a pp-wave background,''
Phys.\ Rev.\ D {\bf 66}, 086004 (2002)
[arXiv:hep-th/0204146].
}

\lref\SpradlinRV{
M.~Spradlin and A.~Volovich,
``Superstring interactions in a pp-wave background. II,''
[arXiv:hep-th/0206073].
}

\lref\PankiewiczGS{
A.~Pankiewicz,
``More comments on superstring interactions in the pp-wave background,''
JHEP {\bf 0209}, 056 (2002)
[arXiv:hep-th/0208209].
}

\lref\PankiewiczTG{
A.~Pankiewicz and B.~J.~Stefanski,
``pp-wave light-cone superstring field theory,''
Nucl.\ Phys.\ B {\bf 657}, 79 (2003)
[arXiv:hep-th/0210246].
}
\lref\ConstableHW{
N.~R.~Constable, D.~Z.~Freedman, M.~Headrick, S.~Minwalla, L.~Motl, A.~Postnikov and W.~Skiba,
``PP-wave string interactions from perturbative Yang-Mills theory,''
JHEP {\bf 0207}, 017 (2002)
[arXiv:hep-th/0205089].
}

\lref\KristjansenBB{
C.~Kristjansen, J.~Plefka, G.~W.~Semenoff and M.~Staudacher,
``A new double-scaling limit of N = 4 super Yang-Mills theory and PP-wave  strings,''
Nucl.\ Phys.\ B {\bf 643}, 3 (2002)
[arXiv:hep-th/0205033].
}

\lref\HuangYT{
M.~x.~Huang,
``String interactions in pp-wave from N = 4 super Yang Mills,''
Phys.\ Rev.\ D {\bf 66}, 105002 (2002)
[arXiv:hep-th/0206248].
}

\lref\PankiewiczPG{
A.~Pankiewicz,
``Strings in plane wave backgrounds,''
Fortsch.\ Phys.\  {\bf 51}, 1139 (2003)
[arXiv:hep-th/0307027].
}

\lref\rustse{
J.~G.~Russo and A.~A.~Tseytlin,
``On solvable models of type IIB superstring in NS-NS and R-R plane wave  
backgrounds,''
JHEP {\bf 0204}, 021 (2002)
[arXiv:hep-th/0202179].
}

\lref\penrose{
R.~Penrose,
``Any spacetime has a plane wave as a limit,''
Differential Geometry and Relativity, Reidel, Dordrecht, 1976,
pp. 271-275.
}

\lref\gms{
J.~Gomis, L.~Motl and A.~Strominger,
``pp-wave / CFT(2) duality,''
[arXiv:hep-th/0206166].
}

\lref\ParnachevQS{
A.~Parnachev and D.~A.~Sahakyan,
`Penrose limit and string quantization in AdS(3) x S**3,''
JHEP {\bf 0206}, 035 (2002)
[arXiv:hep-th/0205015].
}

\lref\HikidaIN{
Y.~Hikida and Y.~Sugawara,
``Superstrings on PP-wave backgrounds and symmetric orbifolds,''
JHEP {\bf 0206}, 037 (2002)
[arXiv:hep-th/0205200].
}

\lref\LuninFW{
O.~Lunin and S.~D.~Mathur,
``Rotating deformations of AdS(3) x S**3, the orbifold CFT and strings in  the pp-wave limit,''
Nucl.\ Phys.\ B {\bf 642}, 91 (2002)
[arXiv:hep-th/0206107].
}

\lref\KazamaQP{
Y.~Kazama and H.~Suzuki,
``New N=2 Superconformal Field Theories And Superstring Compactification,''
Nucl.\ Phys.\ B {\bf 321}, 232 (1989).
}

\lref\GiveonNS{
A.~Giveon, D.~Kutasov and N.~Seiberg,
``Comments on string theory on AdS(3),''
Adv.\ Theor.\ Math.\ Phys.\  {\bf 2}, 733 (1998)
[arXiv:hep-th/9806194].
}

\lref\BerensteinGJ{
D.~Berenstein and R.~G.~Leigh,
``Spacetime supersymmetry in AdS(3) backgrounds,''
Phys.\ Lett.\ B {\bf 458}, 297 (1999)
[arXiv:hep-th/9904040].
}

\lref\GiveonJG{
A.~Giveon and M.~Rocek,
``Supersymmetric string vacua on AdS(3) x N,''
JHEP {\bf 9904}, 019 (1999)
[arXiv:hep-th/9904024].
}

\lref\GubserTV{
S.~S.~Gubser, I.~R.~Klebanov and A.~M.~Polyakov,
``A semi-classical limit of the gauge/string correspondence,''
Nucl.\ Phys.\ B {\bf 636}, 99 (2002)
[arXiv:hep-th/0204051].
}

\lref\deVegaMV{
H.~J.~de Vega and I.~L.~Egusquiza,
``Planetoid String Solutions in 3 + 1 Axisymmetric Spacetimes,''
Phys.\ Rev.\ D {\bf 54}, 7513 (1996)
[arXiv:hep-th/9607056].
}

\lref\GepnerHR{
D.~Gepner and Z.~a.~Qiu,
``Modular Invariant Partition Functions For Parafermionic Field Theories,''
Nucl.\ Phys.\ B {\bf 285}, 423 (1987).
}

\lref\SpradlinXC{
M.~Spradlin and A.~Volovich,
``Light-Cone String Field Theory in a Plane Wave Background,''
[arXiv:hep-th/0310033].
}

\lref\FrolovQC{
S.~Frolov and A.~A.~Tseytlin,
``Multi-spin string solutions in AdS(5) x S**5,''
Nucl.\ Phys.\ B {\bf 668}, 77 (2003)
[arXiv:hep-th/0304255].
}

\lref\FrolovTU{
S.~Frolov and A.~A.~Tseytlin,
``Quantizing three-spin string solution in AdS(5) x S**5,''
JHEP {\bf 0307}, 016 (2003)
[arXiv:hep-th/0306130].
}

\lref\MateosDE{
D.~Mateos, T.~Mateos and P.~K.~Townsend,
``Supersymmetry of tensionless rotating strings in AdS(5) x S**5, and nearly-BPS operators,''
[arXiv:hep-th/0309114].
}

\lref\BonelliZU{
G.~Bonelli,
``On the covariant quantization of tensionless bosonic strings in AdS
spacetime,''
JHEP {\bf 0311}, 028 (2003)
[arXiv:hep-th/0309222].
}

\lref\SagnottiQA{
A.~Sagnotti and M.~Tsulaia,
``On higher spins and the tensionless limit of string theory,''
arXiv:hep-th/0311257.
}

\lref\KiritsisKZ{
E.~Kiritsis and B.~Pioline,
``Strings in homogeneous gravitational waves and null holography,''
JHEP {\bf 0208}, 048 (2002)
[arXiv:hep-th/0204004].
}

\lref\DAppollonioDR{
G.~D'Appollonio and E.~Kiritsis,
``String interactions in gravitational wave backgrounds,''
[arXiv:hep-th/0305081].
}

\lref\GavaXB{
E.~Gava and K.~S.~Narain,
``Proving the pp-wave / CFT(2) duality,''
JHEP {\bf 0212}, 023 (2002)
[arXiv:hep-th/0208081].
}

\lref\BeisertTN{
N.~Beisert,
``BMN operators and superconformal symmetry,''
Nucl.\ Phys.\ B {\bf 659}, 79 (2003)
[arXiv:hep-th/0211032].
}

\lref\BeisertTE{
N.~Beisert, M.~Bianchi, J.~F.~Morales and H.~Samtleben,
``On the spectrum of AdS/CFT beyond supergravity,''
[arXiv:hep-th/0310292].
}                

\lref\KutasovZH{
D.~Kutasov, F.~Larsen and R.~G.~Leigh,
``String theory in magnetic monopole backgrounds,''
Nucl.\ Phys.\ B {\bf 550}, 183 (1999)
[arXiv:hep-th/9812027].
}

\lref\ElitzurMM{
S.~Elitzur, O.~Feinerman, A.~Giveon and D.~Tsabar,
``String theory on AdS(3) x S(3) x S(3) x S(1),''
Phys.\ Lett.\ B {\bf 449}, 180 (1999)
[arXiv:hep-th/9811245].
}

\lref\DharFI{
A.~Dhar, G.~Mandal and S.~R.~Wadia,
``String bits in small radius AdS and weakly coupled N = 4 super  Yang-Mills theory. I,''
[arXiv:hep-th/0304062].
}

\lref\ClarkWK{
A.~Clark, A.~Karch, P.~Kovtun and D.~Yamada,
``Construction of bosonic string theory on infinitely curved anti-de  Sitter space,''
Phys.\ Rev.\ D {\bf 68}, 066011 (2003)
[arXiv:hep-th/0304107].
}

\lref\deMedeirosHR{
P.~de Medeiros and S.~P.~Kumar,
``Spacetime Virasoro algebra from strings on zero radius AdS(3),''
[arXiv:hep-th/0310040].
}

\lref\CallanXR{
C.~G.~Callan, H.~K.~Lee, T.~McLoughlin, J.~H.~Schwarz, I.~Swanson and X.~Wu,
``Quantizing string theory in AdS(5) x S**5: Beyond the pp-wave,''
Nucl.\ Phys.\ B {\bf 673}, 3 (2003)
[arXiv:hep-th/0307032].
}

\lref\ParnachevKK{
A.~Parnachev and A.~V.~Ryzhov,
``Strings in the near plane wave background and AdS/CFT,''
JHEP {\bf 0210}, 066 (2002)
[arXiv:hep-th/0208010].
}

\lref\PlefkaNB{
J.~C.~Plefka,
``Lectures on the plane-wave string / gauge theory duality,''
[arXiv:hep-th/0307101].
}

\lref\SadriPR{
D.~Sadri and M.~M.~Sheikh-Jabbari,
``The plane-wave / super Yang-Mills duality,''
[arXiv:hep-th/0310119].
}

\lref\TseytlinII{
A.~A.~Tseytlin,
``Spinning strings and AdS/CFT duality,''
[arXiv:hep-th/0311139].
}

\lref\PakmanCU{
A.~Pakman,
``Unitarity of supersymmetric SL(2,R)/U(1) and no-ghost theorem for  fermionic
strings in AdS(3) x N,''
JHEP {\bf 0301}, 077 (2003)
[arXiv:hep-th/0301110].
}
    
\lref\PakmanKH{
A.~Pakman,
``BRST quantization of string theory in AdS(3),''
JHEP {\bf 0306}, 053 (2003)
[arXiv:hep-th/0304230].
}

\lref\GiveonKU{
A.~Giveon and A.~Pakman,
``More on superstrings in AdS(3) x N,''
JHEP {\bf 0303}, 056 (2003)
[arXiv:hep-th/0302217].
}

\lref\LarsenYW{
A.~L.~Larsen and N.~Sanchez,
``Quantum coherent string states in AdS(3) and SL(2,R) WZWN model,''
Phys.\ Rev.\ D {\bf 62}, 046003 (2000)
[arXiv:hep-th/0001180].
}

\Title{\vbox{\baselineskip12pt\hbox{hep-th/0312017}\hbox{}}}
{\vbox{\centerline {Strings on Plane Waves and $AdS \times S$}
}}

\centerline{John Son}
\bigskip\centerline{Department of Physics}
\centerline{Harvard University} 
\centerline{Cambridge, MA 02138}
\bigskip\centerline{{\tt json@pauli.harvard.edu}}

\vskip .3in \centerline{\bf Abstract}
We consider the RNS superstrings in $AdS_3 \times S^3 \times {\cal M}$, 
where $\cal M$ may be $K3$ or $T^4$, based on $SL(2,R)$ and $SU(2)$ WZW 
models.   We construct the physical states and calculate the spectrum.
A subsector of this theory describes strings propagating
in the six dimensional plane wave obtained by the Penrose limit of
$AdS_3 \times S^3 \times {\cal M}$.  We reproduce the plane wave spectrum 
by taking $J$ and the radius to infinity.
We show that the plane wave spectrum actually coincides with the 
large $J$ spectrum at fixed radius, i.e.~in $AdS_3 \times S^3$.
Relation to some recent topics of interest such as the Frolov-Tseytlin
string and strings with critical tension or in zero radius $AdS$ are discussed.

\smallskip 
\Date{}

\listtoc

\writetoc

\newsec{Introduction}

Understanding the duality between gauge theories and string theories has
been a topic
of great interest, ever since \refs{\thooft}.  The $AdS$/CFT correspondence  
\refs{\MaldacenaRE , \WittenQJ , \GubserBC , \AharonyTI}
has provided a wealth of examples in which string theories compactified
on spacetimes that include an $AdS$ factor are dual to field theories on
the boundary of the $AdS$.  However, progress has been limited due to the fact
these backgrounds involve R-R fields, and solving for the string spectrum in 
such spacetimes remains an outstanding problem.  
Lacking an exact string theory,
the low-energy supergravity approximation has been used for most part.

Recently an exciting set of new dualities was obtained by Berenstein,
Maldacena, and Nastase \refs{\bmn}.  
By taking a limit of $AdS_5 \times S^5$ 
in which the geometry becomes that of a plane wave, one obtains a background
that allows for exact string quantization, in Green-Schwarz formalism
\refs{\MetsaevBJ}.  The limiting procedure involves taking the radius of
$AdS_5 \times S^5$ to infinity and is an example of Penrose's
limit \refs{\penrose}.
At the same time, on the CFT side one focuses on those states with 
large conformal weight and R-charge: \  $\Delta, J \to \infty$  
as $R^2$, but with finite $\Delta -J$.  In this way each $AdS$/CFT 
duality gives rise to a plane wave/CFT duality, in which one may go beyond
the supergravity approximation.  Specifically, BMN was able to reproduce,
from the CFT point of view, some of the {\it stringy} excitations in the 
plane wave.
This represents remarkable progress towards establishing the correspondence
between a fully string theoretic description of gravity on $AdS$ and the CFT
on the boundary.  
 
Furthermore, it has been shown that some 
physical quantities of interest 
may be computed perturbatively on both sides of the BMN 
correspondence \refs{\KristjansenBB,\BerensteinSA,\ConstableHW,\SpradlinAR}.
This differs from  $AdS$/CFT, in which the duality relates the weak 
coupling physics on one side to the strong coupling physics on the other.  
This development has led to an intense level of activity\foot{See, 
for example, \refs{\KiemXN, \LeeRM, \SpradlinRV, \PankiewiczGS, \PankiewiczTG
\HuangYT, \BeisertBB, \ConstableVQ, \LeeVZ}, 
and \refs{\PankiewiczPG, \PlefkaNB, \SpradlinXC, \SadriPR} 
for reviews.}
which has resulted in significant understanding
of both gauge theory and string field theory.  

For these reasons string theory on plane waves that arise as Penrose 
limits of $AdS \times S$ has
emerged as a topic of great importance.  
However, the GS formulation of superstrings is technically cumbersome
and much insight would be gained from an example of an exact CFT 
description of 
string propagation on a plane wave.  Happily, such an example exists:  the 
plane wave obtained via the Penrose limit of 
$\ads \times S^3$ with a purely NS-NS field strength.  One of the goals of 
this paper is to study superstrings in this background using CFT techniques.
For earlier work on the $\ads \times S^3$ plane wave, see
\refs{\rustse, \ParnachevQS, \HikidaIN, \KiritsisKZ, \GavaXB, \LuninFW}.
   
Actually, the $\ads \times S^3$ plane wave with NS background is special 
for another reason--string theory is solvable even 
before the Penrose limit is taken!    
The CFT on the string worldsheet is    
given by the $\SL$ and $\SU$ WZW models with the level of the current
algebras
determined by the radius of $\ads \times S^3$.  
With the recent understanding of the 
$\SL$ WZW model \refs{\moo ,\mos ,\mot}, 
the string spectrum in $\ads$ has been found and
correlation functions have been calculated.  
The solvability of string theory on $\ads \times S^3$ allows us to 
view string theory on the six dimensional plane wave
as one of its subsectors.  
This is similar in spirit to how the ${\cal N}=4$ SYM 
theory is studied in the ten dimensional BMN duality, in that
one does not in anyway
change the theory while trying to study the correspondence.  
Rather, one restricts  focus onto a particular subclass of 
operators, such as the (nearly) chiral operators, 
for which it is possible to say something about the dual objects
in the string side.

In the same manner, we begin with the full string spectrum on 
$\ads \times S^3 \times {\cal M}$ at arbitrary values of the level
$k$ and angular momentum $J$ on $S^3$.  As we take $k, J \to \infty$,
we can ``see'' how the Hilbert space breaks apart, and a subspace
arising in this limit corresponds to the plane wave Hilbert space.
The spectral flow symmetry of the
$\SL$ WZW model found in \refs{\moo} once again will play a key role in
this discussion\foot{Previous work on the plane wave limit of 
$\ads \times S^3$ 
either did not address the issue of spectral flow, or discussed it 
as a symmetry of the WZW model based on the extended Heisenberg group, i.e.
after the Penrose limit was taken.}.

Moreover, since our treatment is fully string theoretic from the start, 
or in other words valid for arbitrary values of the radius, we can 
attempt to address the following important question:  What can string
theory on plane waves tell us about string theory on $AdS \times S$?  Even
though it is believed that the former represents a great simplification
of the latter
(the plane wave is, after all, just the first term in an $R^{-2}$ expansion 
of $AdS \times S$), we find 
some strong evidence that in fact some aspects of string theory in the plane 
wave could be trusted away from the strict $R^2 \to \infty$ limit.  
Specifically, we will
show that the large $J$ spectrum of strings on $\ads \times S^3$ with NS 
background at finite $R^2$ coincides with the plane wave spectrum, found in
\refs{\bmn, \rustse, \ParnachevQS, \HikidaIN}.  This is rather surprising since
the spacetime geometry in each case is drastically different.  
Our result provides an explicit and compelling evidence in support of some 
of the recent 
ideas \refs{\FrolovQC, \FrolovTU, \BeisertTN,\BeisertTE, \TseytlinII}
 about extrapolating the semiclassical relationship between energy
and spins in $AdS_5 \times S^5$ down to the stringy regime.

This paper is organized as follows.  We begin by briefly reviewing 
in section 2
the relevant aspects of string theory on $\ads \times S^3 \times {\cal M}$,
where ${\cal M}$ may be $K3$ or $T^4$.  We first review the bosonic case,
then turn to the case of superstrings which will be the subject of our 
focus.  In section 3, we describe the Penrose limit which takes 
$\ads \times S^3$ to the plane wave. 
In section 4, we study
the semi-classical limit of strings that will be
relevant in the plane wave limit.
We do this by computing the Nambu action of a string near the origin of 
$\ads$ and moving with high angular momentum on a great circle of $S^3$.  
This is the six dimensional analog of the particle trajectory used by BMN 
to obtain
the ten dimensional plane wave from $AdS_5 \times S^5$ \refs{\bmn}.  
The resulting Nambu
action displays the same behavior as what was shown in \refs{\moo}.  Namely,
new representations that do not obey the usual highest weight conditions 
appear.  These representations are obtained from the usual representations
by spectral flow, and it is shown that the amount of spectral flow depends
on the ratio of the angular momentum  to $R^2$.  
Armed with this knowledge, in section 5 we
obtain the exact string spectrum on $\ads \times S^3$, 
valid for arbitrary values of $R^2$ and $J$. 
The plane wave spectrum is reproduced by taking $R^2, J \to \infty$ and
expanding to leading order.
In section 6 we discuss the decoupling of the Hilbert space in 
the Penrose limit.  
In section 7 we discuss what happens when the radius of $\ads \times S^3$
is finite.  Section 8 contains a summary and discussion.  In Appendix A 
we show how the spectral flow number violation rule found in \refs{\mot}
can be understood in terms of angular momentum conservation in the plane wave.

\newsec{String theory on $\ads \times S^3 \times {\cal M}$}
 
We begin with a review of the bosonic string theory in $\ads$ as discussed 
in \refs{\moo}\foot{The list of papers on $\ads$ preceding \refs{\moo} is too
long to be referenced here.  We direct the reader to \refs{\moo} for an 
extensive bibliography.}.
By considering new representations
of $\CSL$ algebra that are obtained by the spectral flow operation, authors of
\refs{\moo} showed that the long-standing problem of an apparent upper limit
on the energy of string states \refs{\BalogJB} was eliminated, 
and that the spectrum consisted
of short strings and long strings \refs{\MaldacenaUZ ,\SeibergXZ}.  
This proposal was verified through 
a one-loop string calculation in \refs{\mos}, and correlation functions 
were computed in \refs{\mot}.   

\subsec{$\SL$ WZW model}

Since $\ads \cong \SL$\foot{We always consider the universal cover of $\SL$, 
obtained by unwrapping the $t$ coordinate.}, the CFT on the string worldsheet
is given by the $\SL$ WZW model.  The action is 
\eqn\wzwact{
S={k \over 8 \pi \apm}\int d^2z \ \Tr (g^{-1}\p g g^{-1} \bar{\p}g) 
+ k\Gamma _{WZ} \;,
}
where $g$ is an element of $\SL$.  The parametrization of $g$ is given by
\eqn\paramsl{
g=e^{i(t+\phi)\sigma^2/2}e^{\rho \sigma^3}e^{i(t-\phi)\sigma^2/2} \;,
}
which corresponds to parametrization  of $\ads$ in the global coordinates
\eqn\adsmet{
ds^2=-\cosh ^2 \rho dt^2+d\rho^2+\sinh^2 \rho d\phi^2 \;.
}  
The level $k$ need not be quantized
since $H^3$ is trivial for $\SL$.  However, when we consider (super)string
theory on $\ads \times S^3$, the level of the two WZW models will turn 
out to be  equal, and owing to the quantization of the level in
the $\SU$ model $k$ is restricted to be an integer.

This model has $\CSL _L \times \CSL _R$
symmetry, generated by two commuting copies of the current algebra
\eqn\comm{\eqalign{
\left[ K^+_m, K^-_n \right] & =  -2K^3_{m+n} + km \delta_{m+n} \cr
\left[ K^3_m, K^{\pm}_n \right] & =  \pm K^{\pm}_{m+n}   \cr
\left[ K^3_m, K^3_n \right] & =  -{k \over 2}m \delta_{m+n}\;. 
}}
The Virasoro generators are given by the Sugawara form 
(we only write the holomophic side from now on)
\eqn\viragen{
L_n={1 \over k-2} \sum_{m=-\infty}^{\infty} : \eta_{ab}J^a_{m}J^b_{n-m}:\;,
}
where $\eta_{ab}$ is the metric on $\SL$ with signature $(+,+,-)$.  
The generators \viragen\ obey the Virasoro algebra with central charge
\eqn\slcc{
c={3 k \over k-2}\;.
}

States of $\SL$ WZW 
model fall into representations of  $\CSL _L \times \CSL _R$ current algebra.
The question is which representations appear in the Hilbert space.
Representations of current algebra can be constructed by
considering representations of the global algebra, generated by the zero modes
of the currents $K_0^{\pm,3}$, to be the primary states annhiliated by 
$K_{m>0}^{\pm,3}$.  Then $K_{m<0}^{\pm,3}$ can be applied to these states,
 generating the representation
of the current algebra.  Hence, the first problem is to find the
right representations of $\SL$.   
In \refs{\moo}, the following was proposed.  The representations of $\SL$ that
appear are
${\cal D}_{\ell}$ and $C_{\ell, \alpha}$, where ${\cal D}_{\ell}$ is the
discrete lowest weight representation
\eqn\glodis{ 
{\cal D}_{\ell} =\{\ket{\ell,n}: n=\ell , \ell+1 , \ell+2 , \dots\} \;,
}
with $K^-_0 \ket{\ell, \ell} =0$.  The representation is labeled by the value 
of the quadratic Casimir
\eqn\quadcas{
\left(\frac12 (K^+_0 K^-_0 + K^-_0 K^+_0)-(K^3_0)^2 \right) \ket{\ell, n}
=-\ell(\ell-1) \ket{\ell, n}\;,
}
and $n$ which is the eigenvalue of $K^3_0$, related to the spacetime energy
by
\eqn\ttan{
E = K^3_0+\bar K^3_0\;.
}
$C_{\ell, \alpha}$ is the continuous representation 
\eqn\glocon{
C_{\ell, \alpha} =\{\ket{\ell,n,\alpha}: n=\alpha 
, \alpha \pm 1 , \alpha \pm2 , \dots\}\;,
}
where without loss of generality $\alpha$ may be restricted to 
$0 \le \alpha <1$.  Unitarity requires $\ell = 1/2+is$ with $s$ real.  This
gives for the quadratic Casimir
\eqn\contcas{
\left(\frac12 (K^+_0 K^-_0 + K^-_0 K^+_0)-(K^3_0)^2 \right) 
\ket{\ell, n,\alpha}
=\left(\frac14 + s^2\right)\ket{\ell, n, \alpha}\;.
}

Now starting with the above representations of $\SL$, representations of
$\CSL$ are generated by applying $K^a_{m<0}$.  The resulting representations
are denoted $\hat{{\cal D}}_{\ell}$ and $\hat{C}_{\ell, \alpha}$.  However,
these are not the only representations of $\CSL$ that may appear.
There are additional representations that are generated by the action
of spectral flow \refs{\moo}
\eqn\wgen{\eqalign{ 
K^3_m &\rightarrow  \tilde{K}^3_m = K^3_m - {k \over 2}w \delta _{m,0}  \cr
K^+_m &\rightarrow  \tilde{K}^+_m = K^+_{m+w}   \cr
K^-_m &\rightarrow  \tilde{K}^-_m = K^-_{m-w}\;,
}}
and the resulting transformation on the Virasoro generators
\eqn\wvir{
\tilde{L}_m = L_m + w K^3_m -{k \over 4} w^2 \delta_{m,0} \;.
}
For each integer valued spectral flow, we generate the representations
$\hat{{\cal D}}^w_{\ell}$ and $\hat{C}^w_{\ell, \alpha}$ from 
$\hat{{\cal D}}_{\ell}$ and $\hat{C}_{\ell, \alpha}$, respectively.

We are now ready to state the Hilbert space of $\SL$ WZW model.  It is given by
\eqn\slhilb{
{\cal H}_{SL}= \oplus_{w=-\infty}^{\infty} 
\left[ \left(\int_{\frac12}^{{k-1 \over 2}} d \ell \ 
\hat{{\cal D}}^w_{\ell} \otimes \hat{{\cal D}}^w_{\ell} \right) \oplus
\left(\int_{\frac12 +i {\bf R}}d \ell 
\int_0^1 d\alpha \ \hat{C}^w_{\ell, \alpha} \otimes
 \hat{C}^w_{\ell, \alpha} \right) \right] \;.
}
The requirement $\frac12 <\ell <{k-1 \over 2}$ in the case of the discrete
representations is actually more restrictive than what is allowed by the 
no-ghost 
theorem of strings in $\ads$ \refs{\EvansQU}, 
which states that $\ell < {k\over 2}$,
but is needed in order for spectral flow 
symmetry to close upon the representations consistent with harmonic
analysis.  For the continuous representations $\ell=\frac12 + is$ with $s$
a real number.  In the context of string theory on $\ads$, 
$s$ is interpreted as the
momentum in the radial direction, as implied by \contcas.
Finally, we note that the representations with negative spectral flow are
the complex conjugates of the representations with positive spectral flow, 
and in particular generate the $\CSL$ representations built from
discrete highest weight representations of $\SL$.

It is clear that  the $\SL$ WZW model contains states of 
negative norm since the metric
on $\ads$ is of indefinite signature.  However, the situation is no worse
than that of strings in flat space.  There, the field in the time 
direction $X^0$ has the wrong sign and states created by this field have
negative norm.  The situation is remedied by the fact that in covariant
quatization (which we are using for $\ads$), Virasoro 
constraints remove these states, and the resulting
physical string spectrum is free of ghosts.
In \refs{\moo}, it was shown that the same is true for strings on $\ads$,
extending the earlier results of \refs{\EvansQU}.  In the following, 
we will consider a critical bosonic string theory on 
$\ads \times {\cal X}$.  We will     
impose the Virasoro constraints on the product Hilbert space
of $\SL$ WZW model and the CFT describing string propagation on ${\cal X}$,
and obtain the physical spectrum.

\subsec{Bosonic strings on $\ads \times {\cal X}$}

We assume that the CFT on ${\cal X}$ is unitary, with central charge
\eqn\boscen{
c_{\cal X}=26-{3k \over k-2}\;.
}
The Virasoro operators are given by the sum of the Virasoro operators for
each CFT, $L_m=L_m^{SL}+L_m^{{\cal X}}$.
Consider a state in the discrete representation of
$\SL$ WZW model, tensored with a state from ${\cal X}$
with conformal weight $h$.  The combined state is labeled as
\eqn\totst{
\ket{\tilde{\ell}, \tilde{n}, \tilde{N}, w, h}\;,
}
where $N$ is the grade\foot{The word grade is used here instead of level,
as the latter will be reserved for $k$ that appears in \wzwact .} of the 
$\CSL$ descendent.  Let us denote this state by 
$\ket \Omega$.  Taking into account the spectral flow relations
\wgen\ and \wvir , the Virasoro constraints are
\eqn\vircons{\eqalign{
(L_0 -1)\ket{\Omega} &= 
\left(-{\tilde{\ell}(\tilde{\ell}-1) \over k-2} +\tilde{N}-w\tilde{n} 
-{kw^2 \over 4}+h-1 \right)
\ket{\Omega} =0 \cr
L_m\ket{\Omega} & =
(\tilde{L}^{SL}_m-w\tilde{K}^3_m+L_m^{{\cal X}})\ket{\Omega}=0 
\;, \ \ \ \ \ \ m \ge 1 \;.
}}
For discrete representations, $\tilde{n}=\tilde{\ell}+q$, 
with $q$ an integer.  
Using this relation with the first equation in \vircons , 
$\tilde{\ell}$ is determined to be
\eqn\detell{
\tilde{\ell}= \frac12 - {k-2 \over 2}w +
\sqrt{\frac14 +(k-2)
\left(N+h-\frac12 w(w+1)-1\right)} \;,
}
where $N$ is the grade as measured by $L_0$, related to $\tilde N$ by
$N = \tilde{N}-wq$.  We impose the level matching condition
$L_0=\bar{L}_0$, and find the spacetime energy \ttan\
\eqn\spted{
E= 1 + 2w + q + \bar{q} 
+\sqrt{1 +4(k-2)\left(N_w+h-\frac12 w(w+1)-1\right)} \;.
}
Note that
the energy is discrete, even though in the $\SL$ WZW model \slhilb\
the spectrum of $\ell$ was continuous.  These states correspond to
the short strings that are trapped inside $\ads$.

For states coming from the continuous representations, we can 
proceed in a similar manner to obtain their spectrum.  The difference in this
case is that $\tilde \ell$ and $\tilde n$ are not related.  The result is
\eqn\sptec{
E={kw \over 2}+{1 \over w}\left({2s^2 + \frac12 \over k-2}
+\tilde{N} +\tilde{\bar{N}}+h+\bar{h}-2\right)\;.
} 
Note that this time the grade is measured by $\tilde{L}_0$.  
The spectrum is continous and $s$ represents the momentum of the string
in the radial direction of $\ads$.  These are the long strings 
\refs{\SeibergXZ , \MaldacenaUZ} that
can approach arbitrarily close to the boundary.  Note that the 
continuous representations with $w=0$ correspond to a tachyon in the
spacetime.  When we consider the supersymmetric case, such states
will be projected out of the spectrum.

In this paper we we will be interested in the case where the internal
space $\cal X$ contains $S^3$.
For this we will need the $\SU$ WZW model, which we turn to next.  We will
briefly describe the Hilbert space of $\SU$ WZW model, in order 
to introduce 
notation and also because as we will see later, the analog of spectral flow
\wgen , \wvir\ in the $\SU$ WZW model will prove to be an useful tool in
studying superstrings in the plane wave.

\subsec{$\SU$ WZW model}

String theory on $S^3$ is described by the $\SU$ WZW model, 
and its Hilbert space can be constructed in a manner similar to what we
described above for $\SL$.  Again, we will restrict our attention to 
the holomorphic sector.

The action takes the same form as \wzwact, but now with $g$ labelling
an element of $\SU$.  The parametrization of the $\SU$ group manifold
is very similar to what was used for $\SL$, eqn.~\paramsl.  
The metric on $S^3$ reads
\eqn\sumet{
ds^2= \cos ^2 \theta d \psi ^2 + d \theta ^2 +\sin^2 \theta  d \varphi^2\;.
}    

The symmetry of $\SU$ WZW model is generated by two copies of the
$\CSU$ current algebra at level $k'$
\eqn\sucalg{\eqalign{
\left[ J^+_m, J^-_n \right] & =  2J^3_{m+n} + k'm \delta_{m+n} \cr
\left[ J^3_m, J^{\pm}_n \right] & =  \pm J^{\pm}_{m+n}   \cr
\left[ J^3_m, J^3_n \right] & =  {k' \over 2}m \delta_{m+n}\;, 
}}
and the Virasoro algebra given by the Sugawara form
\eqn\virasu{
L_n={1 \over k'+2}\sum_{m=-\infty}^{\infty}:\delta_{ab}J^a_m J^b_{n-m}:\;.
} 
The representations of the $\SU$ WZW model are built
from the familiar $\SU$ angular momentum representations $D_j$.  
A state is labeled as
$\ket{j, m, M}$, with 
\eqn\sustate{\eqalign{
L_0 \ket{j, m, M} &= \left({j(j+1) \over k'+2}+M\right)  \ket{j, m, M} \cr
J_0^3 \ket{j, m, M} &= m\ket{j, m, M} \;.
}}
The zero modes of $J^3$ and $\bar{J}^3$ are related to translation
along $\psi$ direction in \sumet :
\eqn\angmom{
-i{\p \over \p \psi} = J_0^3+\bar{J}^3_0 \;.
}
The possible values of $j$ that
may appear are restricted to $0 \le j \le k'/2$, 
in half-integer steps \refs{\GepnerWI}.  The complete Hilbert space
of $\SU$ WZW model is therefore
\eqn\suhilb{
{\cal H}_{SU}=\oplus_{j=0, \frac12 , \dots {k' \over 2}} 
\hat{D}_j \otimes \hat{D}_j \;.
}

\subsec{Superstrings on $\ads \times S^3 \times {\cal M}$}

So far we have discussed the bosonic string theory.  
Our main interest is in the supersymmetric case, and  
in this subsection we will describe the supersymmetric
extension of above discussion.  For simplicity, we will limit our discussion
to the $\SL$ model; the corresponding modifications for the $\SU$ model
is straightforward.  Further details on superstrings on
group manifolds can be found in \refs{\KazamaQP}.  Superstrings on 
$\ads \times S^3$ was also studied in \refs{\IsraelRY}, and the no-ghost 
theorem was proved in \refs{\PakmanCU,\PakmanKH}.

To extend the above results to the case of superstrings in RNS formalism,
we need to introduce free worldsheet fermions 
$\chi^a$ which together with
the total current $K^a$ comprise the WZW supercurrent:
\eqn\supercur{
C^a=\chi^a+\theta K^a\;,
}
with $\theta$ a holomorphic Grassmann variable.  
The OPE's of $K^a$ and $\chi^a$ are
\eqn\opecur{\eqalign{
K^a(z)K^b(w) &\ \sim \ {k \over 2} {\eta^{ab} \over (z-w)^2}
+{i\epsilon^{ab}_{\ \ c} K^c (w) \over z-w} \cr
K^a(z)\chi^b(w) & \ \sim \ {i\epsilon^{ab}_{\ \ c} \chi^c (w) \over z-w} \cr
\chi^a(z)\chi^b(w) & \ \sim \ {k \over 2}{\eta^{ab} \over z-w} \;.
}}
This shows that $K^a$ and $\chi^a$ do not form independent algebras.
By subtracting the fermionic contribution to the total current, we
obtain the bosonic current
\eqn\boscur{
k^a=K^a+{i \over k} \epsilon^a_{\ bc}\chi^b\chi^c \;,
}
which have the OPE's
\eqn\opebos{\eqalign{
k^a(z)k^b(w) &\ \sim \ {k+2 \over 2} {\eta^{ab} \over (z-w)^2}
+{i\epsilon^{ab}_{\ \ c} k^c (w) \over z-w} \cr
k^a(z)\chi^b(w) &\ \sim  \ 0 \;.
}}
Hence the level of the bosonic WZW model is shifted from $k$ to $k+2$.
Similarly, for the supersymmetric $\SU$ WZW model one introduces 
three fermions $\zeta^a$ which
together with $J^a$ form the supercurrent.
The purely bosonic current $j^a$ is defined analogous to \boscur, and
the level of the 
bosonic part is shifted from $k'$ to $k' -2$.
The stress tensor and the Virasoro supercurrent are given by
\eqn\sustress{\eqalign{
T&={1\over k}(\eta_{ab}k^ak^b -  \eta_{ab} \chi^a \p \chi^b)
+{1\over k'}(\delta_{ab}j^aj^b-\delta_{ab} \zeta^a \p \zeta^b)\cr
G&={2 \over k}\left(\eta_{ab} \chi^a k^b 
- {i \over 3k} \epsilon_{abc}\chi^a \chi^b \chi^c \right)
+{2 \over k'} \left(\delta_{ab} \zeta^a j^b 
- {i \over 3k'} \epsilon_{abc}\zeta^a \zeta^b \zeta^c \right)\;.
}}

Criticality of  superstring theory on 
$\ads \times S^3 \times {\cal M}$, where ${\cal M}$ is $K3$ or $T^4$,
requires the central charge to satisfy
\eqn\totcc{
{3(k+2) \over k}+ \frac32+{3(k'-2) \over k'} + \frac32 = 9\;,
}
which relates the levels of the current algebras
\eqn\rellev{
k=k'\;.
}
It is worthwhile to use variables commonly used when discussing 
$AdS$/CFT duality.
In deriving the $\ads /{\rm CFT}_2$ correspondence from D-branes, 
S-duality can be used to transform the D1-D5 system into an NS1-NS5 system.
Taking the near horizon limit, the level of $\SL$ WZW model 
is identified with  $Q_5$,
the number of 5 branes 
(for details see \refs{\GubserBC}, \refs{\MaldacenaBW}).
Hence the bosonic levels of the $\SL$ and $\SU$ WZW models
are $Q_5+2$ and $Q_5-2$, respectively.

The supersymmetric generalization of spectral flow in $\SL$ WZW model
was given in \refs{\ArgurioTB}.  
The spectral flow operation, given by the action of
what was referred to as the ``twist field'' in that work, not only induces
transformation on the $\CSL$ quantum numbers but also on the CFT describing
the internal space.  
Physically, this coupling between the $\SL$ part and the internal CFT
has its roots in the fact that in order for the spacetime
theory to admit supersymmetry, one needs to pair $\chi^3$ with a fermion from
the internal CFT and then bosonize \refs{\GiveonNS, \BerensteinGJ, \GiveonJG}. 
In the case of $\ads \times S^3 \times {\cal M}$ the internal fermion 
is identified with $\zeta^3$ and in the language of \refs{\ArgurioTB} every
time the twist in $\CSL$ is taken there is a corresponding twist in $\CSU$.
 
Thinking of spectral flow as a twist is equivalent to the parafermion
decomposition $\SL \simeq \SL/U(1) \times U(1)$ and 
$\SU \simeq \SU/U(1) \times U(1)$, in the following way.
Introduce free bosons $\phi$ and $\psi$, normalized such that
\eqn\psicor{
\lra {\phi (z) \phi (z')} =  \log (z-z')\;, \ \ \ \ \ 
\lra {\psi (z) \psi (z')} = - \log (z-z')\;.
}
In terms of which  $k^3_0$ and $j^3_0$ can be expressed as
\eqn\psij{
k^3(z)  =-i\sqrt{k \over 2}\p \phi\;, \ \ \ \ \ 
j^3(z)  =-i\sqrt{k' \over 2}\p \psi\;.
}
Throughout this discussion $k$ and $k'$ stand for the bosonic $\SL$ and $\SU$ 
levels, respectively.
Then the bosonic $\SL$ primary field $\Phi_{ln\bar{n}}$ is 
decomposed into a field
of $\SL/U(1)$ times a field in $U(1)$, where the $U(1)$ is generated by $\phi$:
\eqn\slpri{
\Phi_{ln\bar{n}} = e^{in\sqrt{2\over k}\phi +i\bar{n}\sqrt{2\over k}\phi}
\Phi^{SL/U(1)}_{ln\bar{n}} \;.
}   
Similarly, a bosonic $\SU$ primary $\Psi_{jm\bar{m}}$ is written as
\eqn\supri{
\Psi_{jm\bar{m}} = e^{im\sqrt{2\over k'}\phi +i\bar{m}\sqrt{2\over k'}\phi}
\Psi^{SU/U(1)}_{jm\bar{m}} \;.
}   
The fields $\Phi^{SL/U(1)}_{ln\bar{n}}$ are 
$\Psi^{SU/U(1)}_{jm\bar{m}}$ parafermions, with weight
\eqn\hsupara{\eqalign{
h(\Phi^{SL/U(1)}_{ln\bar{n}})= -{l(l-1) \over k-2}+{n^2 \over k}\;, \cr
h(\Psi^{SU/U(1)}_{jm\bar{m}})= {j(j+1) \over k'+2}-{m^2 \over k'}\;,
}}
so that \slpri\ and \supri\ have the expected weights.  
Note that under the shift $n \to n +wk/2$ and $m \to m+wk'/2$,   
the weights of the primary fields change to
\eqn\chpri{\eqalign{
h(\Phi_{ln\bar{n}}) \to
-{l(l-1) \over k-2}-nw -{kw^2 \over 4} \;,\cr
h(\Psi_{jm\bar{m}}) \to
{{j}({j}+1) \over k'+2}+{m}w +{k'w^2 \over 4} \;.
}}
Spectral flow in the supersymmetric theory consists of the above shift
in $n,m$, plus an additional contribution from the fermions 
\refs{\ArgurioTB}, which gives
\eqn\sushft{\eqalign{
h(\Phi_{ln\bar{n}}^w) =
-{l(l-1) \over Q_5}-nw -{Q_5w^2 \over 4} \;,\cr
h(\Psi_{jm\bar{m}}^w) =
{{j}({j}+1) \over Q_5}+{m}w +{Q_5w^2 \over 4} \;.
}}
There is a similar relation on the anti-holomorphic side as well, with
the same $w$. 
Note that the parafermion formalism also provides a convenient way of 
defining the vertex operators
for states belonging to the spectral flowed representations 
\refs{\GepnerHR, \moo, \mot}.
The physical state condition is 
$(L_n-a\delta_{n,0})\ket \Omega =0$ for $n \ge 0$, 
where $a=\frac12$ in the NS sector and
$a=0$ in the R sector, as well as $G_r \ket \Omega =0$ for $r\ge 0$. 
In addition, the analogue of GSO projection is the requirement of mutual 
locality with the supercharges that are constructed by bosonizing
the worldsheet fermions \refs{\ArgurioTB}.

\newsec{Penrose limit of $\ads \times S^3$ with NS background}

In this section we explain the Penrose limit \refs{\penrose} 
of $\ads \times S^3$ that results in the plane wave geometry
\refs{\bmn, \gms}.

The six dimensional plane wave is obtained from 
$\ads \times S^3$ by expanding
around a particular class of geodesics.  These geodesics correspond
to a particle near the center of $\ads$ and moving with very high
angular momentum around a great circle of $S^3$.  For this purpose, we
begin with the spacetime metric
\eqn\smetric{
ds^2 =  R^2(-\cosh ^2 \rho dt^2 + d \rho ^2 + \sinh^2 \rho d\phi ^2+
\cos ^2 \theta d \psi ^2 + d \theta ^2 +\sin^2 \theta  d \varphi^2)
+ds^2_{{\cal M}} 
}
and introduce the coordinates
\eqn\ppcoord{\eqalign{
t &= \mu x^+  \ \ \ \cr
\psi &=\mu x^+ -{x^-\over \mu R^2} \;.
}}
Rescaling $\rho = {r/ R}$,  $\theta= {y / R}$, the metric is expanded
around $\rho =\theta=0$ by taking the limit $R \to \infty$.  This results
in the six dimensional plane wave
\eqn\ppmet{
ds^2=-2dx^+dx^- -\mu^2(r^2+y^2)dx^+dx^+ + dr^2 + r^2 d \phi^2
+d y^2 + y^2 d {\varphi}^2 + ds^2_{{\cal M}} \;.
}  

String spectrum in this background with NS three form field strength
was found by quantizing the light cone action in 
\refs{\bmn, \rustse, \HikidaIN}.  
For our purposes we will find it convenient to take the  
light cone Hamiltonian as given
in \refs{\ParnachevQS}, adapted to the conventions of this paper and 
supersymmetrized,
\eqn\lighth{
H_{lc}=p^-=\mu(2+q+\bar q)
+{N+\bar N+ h^{{\cal M}}+\bar{h}^{{\cal M}} -1\over p^+ \apm}\;.
}
This applies to the NS-NS sector, and the last term needs to be appropriately
changed for the R sector.  The quantities appearing in this 
expression have the following physical interpretation.  
$N$ is the total grade coming from the excitations along
the pp-wave.  $h^{\cal M}$ is the weight of the state coming from the CFT 
on $\cal M$.  Finally, $q$ is the net number of times the spacetime light 
cone energy raising and lowering operators have been applied to the
ground state.   The ground state in question may or may not be physical, 
i.e.~we are referring to the ground state before the GSO projection.
We have chosen the letter $q$ to denote this number because as we shall
see the physical meaning of this quantity is the same as the $q$
we used in labelling the current algebra representations, see the remark
below \vircons.  There are corresponding contributions from the 
anti-holomorphic side to \lighth, subject to the constraint that the net
momentum along the worldsheet vanishes,
\eqn\gradematch{
N+h=\bar N+\bar h\;.
} 
      
The lightcone variables $p^-$ and $p^+$ are related to observables measured
in the global coordinates \smetric\ by 
\eqn\rellc{\eqalign{
p^- & =  i\p_{x^+} = \mu(E-J) \cr
p^+ & =  i\p_{x^-} = {J\over \mu R^2}\;.
}}
$E$ is the spacetime energy and $J$ is the angular momentum around the $\psi$
direction in $S^3$.  Our choice of basis in labeling the $\SU$ representations
\sustate\ corresponds to diagonalizing the action of rotation in $\psi$,  
hence $J$ is given by $J^3_0 + \bar{J}^3_0$.

The radius of $\ads$ and $S^3$ is related to $Q_5$ by $R^2=\apm Q_5$, so the
second equation in \rellc \ is equivalent to
\eqn\pplus{
\mu p^+ \apm = {J \over Q_5}.
}
Hence the string spectrum in the NS-NS sector is
\eqn\delj{
E -J = 2+q+\bar q +{Q_5 \over J}(N+\bar N -1) 
+ {Q_5 \over J}(h^{{\cal M}}+\bar{h}^{{\cal M}}) \;,
}
with the condition \gradematch.

We make a few comments about the brane charges.  Note that $Q_1$, 
the number of 1branes, actually never appears in
any of the formulas\foot{This is a feature of the NS1-NS5 description
\refs{\MaldacenaBW}.}.
But it should be kept in mind that $Q_1$ is being taken to infinity as well.
As explained in \refs{\gms}, the plane wave limit can be
described in terms of the brane charges by taking $Q_1$, 
$Q_5 \to \infty$, with fixed $Q_1/Q_5$.  The scaling used to obtain the 
plane wave requires that finite energy excitations of 
the resulting geometry have $\Delta$, $J \to \infty$ as $\sqrt{Q_1Q_5}$, with
finite $\Delta - J$.  Since $Q_1 \propto Q_5$, this actually implies that
$\Delta$, $J \to \infty$ as $Q_5 \sim k$, the level of the current algebra.
We could have seen this directly from the fact that $J/R^2$ is held fixed
as the limit $R^2 \to \infty$ is taken, but then it would not
be clear that $Q_1$ is scaled to infinity as well.   
Also note that  in the case of  $Q_5 =1$, 
due to the aforementioned shift in the level of the bosonic
WZW model the bosonic $\SU$ part has a negative level.  This is in 
conflict with the well-known result that the $\SU$ level must be
a non-negative integer.  We will return to the issue of $Q_5=1$
later in section 7.

\newsec{Nambu action near the origin of $\ads \times S^3$}

One of the things we want to understand is how the string
spectrum on $\ads \times S^3 \times {\cal M}$ reduces to \delj\ in the limit
$Q_5, J \to \infty$.  In order to answer this question we must first
understand how \delj\ takes into account the spectral flow parameter $w$.
In this section we explain the physical significance of spectral flow in the 
plane wave.

The plane wave limit described above is essentially a semi-classical
expansion about $\ads \times S^3$, combined with the unusual procedure of
boosting to infinite (angular) momentum.
Indeed, the large $k$ limit in WZW models corresponds to the semi-classical
limit, since the WZW action is proportional to $k$.  Motivated by these
concerns we will consider the Nambu action, upto quadratic order in the
fields, of a string moving moving near 
$\rho \sim \theta \sim 0$ of $\ads \times S^3$.  When $J$ is taken to be 
large, of order $Q_5$, the resulting action displays spectral asymmetry
which is then related to spectral flow \refs{\moo}.

The Nambu action is given by
\eqn\nambu{
S={1\over 2\pi \apm}\int d\tau d\sigma (\sqrt{|g|}-\epsilon_{ab}
B_{\mu \nu} \p_a X^{\mu}\p_b X^{\nu})
}
with $g$ the induced metric and $B_{\mu \nu}$ the NS-NS two form.  
The non-zero components of the $B$ field are
\eqn\bns{
B_{t \phi} =\frac14 \apm Q_5 \cosh 2 \rho \;, \ \ 
B_{\psi \varphi} = \frac14 \apm Q_5 \cos 2 \theta \;.
}
We will consider a string located at small values of $\rho$ and $\theta$, 
and moving along the $\psi$ direction.  Since we will be interested in states
with fixed angular momentum around $\psi$, we take as our classical solution
$\psi (\tau, \sigma)=\psi (\tau)$.  This corresponds to a string collapsed
to a point and rotating around a great circle\foot{The importance of studying
such solutions were pointed out in \refs{\GubserTV, \deVegaMV}.}.    
The components of the induced metric 
$g_{ab}=G_{\mu \nu} \p_a X^{\mu} \p_b X^{\nu}$ are, in the gauge $t =\tau$,
\eqn\indmet{\eqalign{
g_{00}& = \apm Q_5(-(1+(X^a)^2) + \p_0 X^a \p_0 X^a +
(1-(Y^a)^2)(\p_0{\psi})^2+  \p_0 Y^a \p_0 Y^a)  \cr
g_{01}& = \apm Q_5(\p_0 X^a \p_1 X^a +\p_0 Y^a \p_1 Y^a) \cr
g_{11}& = \apm Q_5(\p_1 X^a \p_1 X^a +\p_1 Y^a \p_1 Y^a) \;,
}
}
where $X^1 +i X^2=\rho e^{i\phi}$,
and $Y^1+iY^2=\theta e^{i\varphi}$.
The coupling to B field simplifies in this gauge to
\eqn\bcoup{
-{Q_5 \over 2 \pi}\int d\tau d\sigma (\rho ^2 \p_1 \phi -\theta ^2 \p_0{\psi}
\p_1 \varphi)\;,
}
where we have used the fact that $\psi$ has no dependence on $\sigma$,
\eqn\intbp{
\int d\tau d\sigma \p_0{\psi}\p_1 \varphi
=\int d\tau d\sigma \p_1(\p_0{\psi} \varphi) =0\;.
}
The resulting action \nambu\ shows that $\psi$ is a cyclic coordinate.
Hence, the conjugate momentum $J_0={\p L \over \p (\p_0{\psi})}$ 
is constant and it is advantageous to perform a Legendre transformation
for $\psi$.  The resulting Routhian,
\eqn\routhian{
R(X^a, Y^a ;J_0)=L-J_0\p_0{\psi}\;,
} 
is then the Lagrangian that describes the dynamics of $X^a$ and $Y^a$,
while treating $J_0$ as a constant of motion.  The subscript 0 is added to 
$J$ here to indicate that it is the angular momentum of the ground state,
because we are discussing the point particle limit.
Taking $J_0$ to be large, of
order $Q_5$, the action for 
$X^a$ and $Y^a$ upto quadratic order in the fields is found to be
\eqn\expnd{\eqalign{
S=  {J_0 \over 2 \pi}\int d^2   \sigma & \left[
 1-\frac12|\p_0 \Theta|^2+
\frac12 {1 \over A^2}|(\p_1-iA)\Theta|^2 \right. \cr 
& \left. -\frac12|\p_0 \Phi|^2+
\frac12 {1 \over A^2}|(\p_1-iA)\Phi|^2 \right]\;,
}}
where $A=J_0/Q_5$, and  $X^1+iX^2=\Phi$, $Y^1+iY^2=\Theta$.  
We see that $\Phi$ and
$\Theta$ are two massless charged scalar fields on $R \times S^1$, coupled to
a constant gauge field $A_a=A \delta_{a,1}$.  As shown by Maldacena and 
Ooguri in  \refs{\moo}, this implies that if $A$ is not an integer, the states
of $\Phi$ and $\Theta$ belong to the discrete representations with 
spectral flow number $w$ equal
to the integer part of $A$.  Let us explain how this arises.  The solution
to the equation of motion that follows from \expnd\ is
\eqn\solom{\eqalign{
\Phi &= \sum_n \left(a^{\dag}_n e^{i(n-A)(\tau/A +\sigma)}
+b_n e^{-i(n-A)(\tau/A -\sigma)}\right) {e^{iA\sigma} \over n-A} \cr
\Theta &= \sum_n \left(c^{\dag}_n e^{i(n-A)(\tau/A +\sigma)}
+d_n e^{-i(n-A)(\tau/A -\sigma)}\right) {e^{iA\sigma} \over n-A} \;.
}}
Canonical quantization gives for the commutation relations
\eqn\cancom{\eqalign{
[a_n , a^{\dag}_m] \ & \sim \ (n-A)\delta_{n, m} \;, \ \ \ \
[b_n , b^{\dag}_m] \ \sim \ (n-A)\delta_{n, m} \cr
[c_n , c^{\dag}_m] \ & \sim \ (n-A)\delta_{n, m}\;, \ \ \ \
[d_n , d^{\dag}_m] \ \sim \ (n-A)\delta_{n, m}\;.
}}  
Hence, for $n>A$, $a^{\dag}_n$ is the creation operator while 
for $n<A$, $a_n$ should be thought of as the creation operator.  Similar
comments apply to the other sets of operators.  The holomorphic currents 
constructed from   $\Phi$ and $\Theta$ are
\eqn\concur{\eqalign{
K^{+} & \sim \ -iQ_5\sum_n a_n e^{-in(\tau/A +\sigma)} \cr
K^{-} & \sim \ iQ_5\sum_n a^{\dag}_n e^{in(\tau/A +\sigma)} \cr
J^{+} & \sim \ -iQ_5\sum_n c_n e^{-in(\tau/A +\sigma)} \cr
J^{-} & \sim \ iQ_5\sum_n c^{\dag}_n e^{in(\tau/A +\sigma)} \;. 
}}
Each current may be mode expanded and using \cancom\ the vacuum obeys
\eqn\vacrel{\eqalign{
n>A: \ \ \ \ J^+_n \ket 0 &= 0 \;, \ \ \ \ K^+_n \ket 0 = 0 \;, \cr
n>-A: \ \ \ \ J^-_n \ket 0 & =0 \;, \ \ \ \ K^-_n \ket 0 = 0 \;.
}}
Notice that this is different from the familiar highest weight conditions,
which state that, for example, $K^+_{n>0}$ should annihilate the vacuum.
The highest weight conditions can be restored by the transformation
\eqn\resttr{
K^{\pm}_n=\tilde{K}^{\pm}_{n\mp w} \;, \ \ \ \ \ \ 
J^{\pm}_n=\tilde{J}^{\pm}_{n\mp w} \;,
}
with $w$ an integer satisfying $w<A<w+1$.  With respect to $\tilde{K}$ and
$\tilde{J}$, the states created from $\ket 0$  fill out the conventional 
highest weight representations.  
This shows that for $J_0$ not a multiple of $Q_5$, the states are 
in the discrete representations with spectral flow number equal to the integer
part of $J_0/Q_5$.

On the other hand, when $J_0/Q_5$ is an integer, the $\SL$ part of the state
is in the continuous representation with spectral flow number $J_0/Q_5$ 
\refs{\moo}.  

The fact that spectral flow is necessary when $J_0$ is comparable to $Q_5$
should not be too surprising.  
In fact, the role of spectral flow is precisely to
resolve the apparent conflict between the upper limit on $\SL$ spin of
the discrete representations \slhilb\
 and the freedom to have arbitrarily high angular 
momentum on $S^3$.  More generally, for spacetimes of the form 
$\ads \times {\cal N}$, the analysis of \refs{\moo}
shows that the amount
of spectral flow is determined by ratio of the 
conformal weight $h$ coming from the operator of the  $\cal N$ CFT    
to the $\SL$ level $k$,
\eqn\genspec{
w < \sqrt{4h \over k} <w+1 \;.
}
For the case at hand, we see that $4h$ can be approximated as
${J^2_0 / k}$ and
using $k\sim Q_5$ this reproduces what we found above.
     
What is surprising, however, is that \expnd\ and the arguments that follow it
imply that spectral flow should also be taken in the $\SU$ theory, with the
same amount as the $\SL$ part.  To be sure,
this is not to suggest that the Hilbert space of $\SU$ WZW model needs to be
enlarged to include spectral flowed representations, 
similar to what was done in the case of $\SL$ model.
Whereas the $\CSL$ representations generated
by spectral flow  are new and distinct from the conventional
representations, this is not true in the case of $\CSU$ representations. 
But as reviewed in section 2 supersymmetry requires that spectral flow is
taken in both WZW models.  Due to the high number of supersymmetries 
possible on this 
background\foot{String theory on $\ads \times \cal N$ generically 
has $N=2$ spacetime supersymmetry if $\cal N$ has 
an affine $U(1)$ symmetry and the coset ${\cal N}/U(1)$ admits a 
$N=2$ superconformal algebra.  In the case ${\cal N}=S^3 \times \cal M$, 
supersymmetry is enhanced to $N=4$ 
\refs{\BerensteinGJ, \GiveonJG, \GiveonKU}.} 
it is not unreasonable to think that
this peculiar feature of the supersymmetric theory manifests itself in
the purely bosonic analysis presented here.
Additionally, note that
the action of spectral flow on the angular momentum generator,
\eqn\specang{
J^3_0 \to J^3_0 +{wk \over 2},
}
has the right form to be useful in keeping track of states with $J \sim k$
while $k$ is taken to infinity.  This feature makes it worthwhile to
introduce spectral flow in the $\SU$ WZW model\foot{See 
\refs{\DAppollonioDR} for an interesting application of 
spectral flow in the $\SU$ WZW model.}.
In the next section, we will use this
idea to obtain the large $J$ spectrum of superstrings on  
$\ads \times S^3$.

\newsec{The plane wave spectrum}

We now turn to explaining how the plane wave spectrum arises from the 
exact $\ads \times S^3$ results.  
The discussion will be limited to the NS sector, as the R sector can be 
obtained by similar methods, with the additional use of the spin fields.

\subsec{Short strings}

We start with the discrete $w=0$ states, the holomorphic side of which
is labeled by the quantum numbers 
\eqn\unfstate{
\ket{\ell, n, N}\otimes\ket{j, m, M}\otimes \ket {h^{\cal M}}\;.
}
The notation in labeling the $\CSL \times \CSU$ part of the state is the same
as what was used in section 2, and $h^{\cal M}$ is the conformal weight
coming from the CFT on ${\cal M}$. 
In order for \unfstate\ to be physical,
it must satisfy
\eqn\unfsat{
-{\ell (\ell -1) \over Q_5} + {j(j+1) \over Q_5}+N+M+h^{\cal M}=\frac12\;.
}
Let us look for the ground state within a given $j$ sector.  First, we note
that the GSO projection \refs{\ArgurioTB}
requires the lowest excitation number to be one half, 
so from \unfsat\ we find $\ell=j+1$.  
Next, we see that the lowest value of energy
(for fixed $j$) is obtained if this one half unit of excitation
comes from the action of $\zeta^+_{-1/2}$ or $\chi^{- }_{-\frac12}$.  
In the first case, the ground state is
\eqn\jgrnd{
\ket {J/2, J/2} \otimes \zeta^+_{-\frac12}\ket{J/2-1, J/2-1} 
\otimes \ket 0 \;, 
}
and in the second,
\eqn\jgrndtwo{
\chi^{- }_{-\frac12}\ket {J/2+1, J/2+1} \otimes \ket{J/2, J/2}
\otimes \ket 0 \;.
}
Combining with an identical state in the anti-holomorphic side, we see that
there is a total of four states that carry angular momentum $J$ and energy
$E=J$, i.e.~the light cone vacuum.  

We will not discuss the Ramond sector in detail, but in order to complete
the discussion of light cone ground states we briefly mention how many
are found in the Ramond sector.
The number of light cone ground
states coming from the Ramond sector depends on whether $\cal M$ is $T^4$ 
or $K3$.
For $T^4$, there are two ground states in the R sector, and one can construct
the usual NS-NS, NS-R, R-NS, R-R sectors to find a total of 16 ground 
states \refs{\KutasovZH}.
When the internal manifold is $K3$, for the purposes of counting ground
states we can think of $T^4/Z_2$ instead.  Then, as explained in 
\refs{\HikidaIN},
the ground states in the NS-R and R-NS sectors are projected out, and
the 16 twisted sectors each give a ground state in the R-R sector.  Thus
there are 24 ground states in all, as expected.

The excited states of $w=0$ representations
are obtained from the lowest weight of $SL(2,R)$ and the highest
weight of $SU(2)$ by the action of negatively
moded generators.  Physical states do not carry excitations along the time
direction.  For example, in the $\SL$ Hilbert space those states satisfying 
the Virasoro conditions can be written 
\eqn\ongnd{ 
\prod_{r=1/2}^\infty({\chi}^+_{-r})^{{N}_r^+}({\chi}^-_{-r})^{N_r^-}
\prod_{n=0}^\infty({k}^+_{-n})^{{N}_{n}^+}
({k}^-_{-n})^{{N}_{n}^-}
\ket{{\ell}, n={\ell}}\;,
}
which has the grade
\eqn\whichhas{
N = \sum_n n(N_n^+ + N_n^-)+\sum_r r(N_r^+ +N_r^-)
}
and $n=\ell+q_{SL}$, with
\eqn\defq{
q_{SL} = \sum_n (N_n^+ - N_n^-) +\sum_r (N_r^+ - N_r^-) \;.
}
Similar relations hold for the $\SU$ part.  Now 
\unfsat\ 
is used to solve for $\ell$, which then gives for the energy 
\eqn\eflim{
E = 1+q_{SL}+\bar{q}_{SL}
+\sqrt{(2j+1)^2 +2Q_5 
(N+\bar{N}+M+\bar{M}+h^{\cal M}+\bar{h}^{\cal M}-1)} \;,
}
with $j$ related to $J$ by $J=2j-q_{SU}-\bar{q}_{SU}$.  Now we take 
the ``Penrose limit'' $Q_5$, $J \to \infty$ with $J/Q_5$ fixed,
and expanding to terms of order one we find
\eqn\efexp{
E-J=2+ q_{SL}+\bar{q}_{SL}+q_{SU}+\bar{q}_{SU}
+{Q_5 \over J}(N+\bar{N}+M+\bar{M}+h^{\cal M}+\bar{h}^{\cal M}-1)\;.
}
Note that the vacuum states considered above corresponds to 
sum of the $q$'s totalling $-2$ and total grade equal to 1.
That the lowest energy state surviving the GSO projection in the NS sector
has a half unit of excitation is similar to what happens in flat space. 
The difference in this case is that
the various raising operators have different charges under $E$ and $J$. 
Note also that the $w=0$ continuous representations are projected out from
the physical spectrum, since for those representations it is impossible to
satisfy the physical state condition unless $N=0$. 
Hence the spectrum is free of tachyons.

Having understood the $w=0$ states, we now turn to the spectral flowed states.
Consider a state in the spectral flowed representation 
of $\CSL \times \CSU$, tensored
with an operator on ${\cal M}$,
\eqn\wopr{
\ket{w,\tilde{\ell}, \tilde{n}, N}\otimes \ket{w, \tilde{j}, \tilde{m}, M}
\otimes \ket{h^{\cal M}}\;.
}
There is a similar state on the anti-holomorphic side.
Using \sushft, the physical
state condition determines $\tilde{\ell}$ to be
\eqn\condl{
2\tilde{\ell}=1 -{Q_5w}+ 
\sqrt{(2\tilde{j}+Q_5w+1)^2+2Q_5 
\left( N +\bar{N}+M +\bar{M}-2w+h^{\cal M}+\bar{h}^{\cal M}-1 \right)} \;,
}
where we have used the second equation in \sushft\ for the weight of
the $\SU$ state.  In this relation
$N$ and $M$ are the grades measured by $L_0$, not $\tilde L_0$, of the 
$\SL$ and $\SU$ model respectively.  Now we can use 
$J=2\tilde{j}+Q_5w-q_{SU}-\bar{q}_{SU}$ to substitute for $\tilde{j}$
in the expression above, and the energy is given by 
\eqn\engiven{
E=2\tilde{\ell}+Q_5w+q_{SL}+\bar{q}_{SL} \;.
}
This result is an exact formula for the energy of a string state in 
$\ads \times S^3 \times \cal M$ with angular momentum $J$ around $S^3$.

Taking the limit $Q_5$, $J\to \infty$
and expanding to terms of order one, 
\eqn\ejlim{
E - J =2+ q_{SL} + \bar{q}_{SL} + q_{SU} + \bar{q}_{SU}+
{Q_5 \over J}(N + \bar{N} + M + \bar{M} -2w -1) 
+{Q_5 \over J}(h^{\cal M}+\bar{h}^{\cal M})\;.
}
The states with $E=J$ again have the form \jgrnd\ or \jgrndtwo, but now there
is a slight difference due to spectral flow.  For example, in the spectral 
flowed analogue of \jgrnd, fermionic generator is given by
$\tilde\zeta ^+_{-\frac12}$,
which has $M=\frac12 +w$ after taking into account the shift in moding
from spectral flow.  This serves to cancel the extra term in \ejlim\
compared to \efexp.  As found in \refs{\ArgurioTB}, the pattern of chiral
states is relatively simple.  Once the $w=0$ chiral states are identified,
spectral flow generates the chiral states with higher R-charge.   
In general a state similar to \ongnd\ in a
spectral flowed representation has $\tilde N$ and $\tilde q_{SL}$ defined
in the same manner as \whichhas\ and \defq, respectively.  They are
related to what appear above as
\eqn\spechas{\eqalign{
& N  =\tilde N -w \tilde q_{SL} \;, \cr
& q_{SL}  = \tilde q_{SL} \;.
}}

In the semiclassical discussion of the previous section we saw that
the amount of spectral flow necessary is determined by the ratio
$J_0/Q_5$, where $J_0$ is the angular momentum of the ground state, i.e.~a
state in the zero grade of a $\CSU$ representation.  In the fully quantum
treatment, $w$ is determined through the inequality 
$\frac12 < \tilde \ell < \frac{Q_5+1}{2}$, which becomes
\eqn\ineq{
w^2 < 
{( 2\tilde j + Q_5 w+1)^2 \over Q_5^2} + 
{2 \over Q_5}(N +\bar{N}+M +\bar{M}-2w+h^{\cal M}+\bar{h}^{\cal M}-1)
< (w+1)^2 \;.
}
It should be remembered that $N$ and $M$ also depend on $w$, through
\spechas\ and an analogous relation for $M$.  In \ineq\ we can 
think of $\tilde j + Q_5 w/2$ as the highest weight of the $\SU$ 
representation from which the current algebra representation is constructed, 
\eqn\suhas{
J_0 = 2\tilde j + Q_5 w \;,
}
and \ineq\ reproduces the semiclassical result found previously.

\subsec{Long strings and the ``missing'' chiral primaries}

Let us now discuss what happens when the inequality in \ineq\ is saturated, 
which
in the semiclassical approximation corresponds to $J_0/Q_5$ becoming an 
integer. 
In this case we know from \refs{\moo} that the state belongs to a continuous
representation of $\CSL$ with spectral flow number $w=J_0/Q_5$, 
i.e.~it is a long string in $\ads$.  Morever, the energy of the solution
changes smoothly in the transition from a short string to a long string 
(and vice versa).    
The continuous representations do not have highest or lowest weights
and for this reason the spectral flowed states are labelled by the 
eigenvalues of $\tilde L_0$. 
The plane wave spectrum of the long strings is therefore 
\eqn\ejlong{
E-J=2+{Q_5\over J}(\tilde N + \tilde{\bar N}+\tilde M + \tilde{\bar M}-2w-1)
+ {Q_5\over J}(h^{\cal M}+\bar{h}^{\cal M})\;.
}
Sometimes it is possible for a long string to have zero light cone 
energy despite the fact that it is massive. 
If $|0,w\rangle$ denotes a state with $E=J$ then
$k^+_{w} |0,w\rangle$ continues to have zero light cone energy because
$k^+_{w}$'s contribution to \ejlong, proportional to $\tilde{N}$, 
vanishes.  The physical mechanism responsible for this phenomenon
is the same as in $\ads$.  Namely, the coupling to the NS three form
cancels the gravitational attraction.
In the context of plane waves supported by NS field strengths 
it has already been observed that 
there are additional zero modes in the spectrum 
\refs{\bmn, \KiritsisKZ, \DAppollonioDR}, 
which can be understood as the statement that states with special 
values of $p^+$
\ --- \ integer multiples of $1/\mu \apm$ \ --- \ 
do not feel the confining potential of the plane wave.  

It is interesting to note that simplifying $\ads \times S^3$ to the plane
wave makes more apparent the presence of long strings in the spectrum.
As we have just stated, some of the long strings correspond to chiral 
primaries in the dual CFT.  It has been appreciated for a while now that
there is a mismatch of chiral primaries in the $\ads/{\rm CFT}_2$ 
correspondence when considering the $\ads$ with a purely NS background
due to the fact that $\ads$ with vanishing R-R fields corresponds to a
``singular'' CFT \refs{\SeibergXZ, \ArgurioTB}.  The mismatch arises
when one tries to compare the spectrum of chiral primary operators in the CFT
to the spectrum of chiral string states based on the discrete 
representations of $\CSL$.   
It was suggested in \refs{\SeibergXZ} that the chiral
primaries that disappear when all the R-R fields are set to zero might be
found among the continuum.  We find explicitly that indeed there are chiral
primaries belonging to the continuous representations.

\newsec{The decomposition of the Hilbert space in the Penrose limit}

We started with a unitary spectrum of string states in 
$\ads \times S^3 \times \cal M$.
This spectrum is obtained from the Hilbert space of the $\SL$ WZW model,
tensored with the Hilbert spaces of the $\SU$ model and CFT on $\cal M$, 
and imposing the Virasoro constraints.  
In obtaining the results of previous section we have restricted our 
focus to a particular subsector of this physical Hilbert space.  
We now address the question of what happens to the remaining states in
the Hilbert space.  We find that the ratios ${J/Q_5}$,
${J^2 /Q_5}$ determine where the state ends up.

As we take the limit $R \to \infty$, we expect that some of the states 
become strings in flat space, some become strings in the plane wave, 
and the rest with divergent $E-J$.  
The spectra in flat space and plane wave should form
 independent, unitary Hilbert spaces.  Presumably, this means that the states 
with divergent $E-J$ should also, but with a different description.  An
example of such states would be those that have high angular momentum along
a different circle on $S^3$.  These states would be related to what we 
considered above by a global rotation on the sphere.

We have found the states on the plane wave.  
Which states correspond to strings in 
flat space?  In any dimension, flat space is obtained from plane wave 
when \refs{\bmn}
\eqn\flatlim{
\mu \apm p^+ <<1 \;.
}
But in our case, $\mu \apm p^+  = J/Q_5$, and we know
that the integer part of $J/Q_5$ is related to the spectral flow
parameter $w$ in the large $J, Q_5$ limit.  
Thus we conclude that the flat space spectrum comes from 
the unflowed short strings in the original $\ads$ theory.
We can indeed check that for $J/Q_5 \to 0$, ${J^2 / Q_5}$  finite,
the  physical state condition for $w=0$ short strings \unfsat\ 
reproduces the mass formula of superstrings in six flat dimensions
times $\cal M$, because the terms in $L_0$ that involve the quadratic Casimirs
become $p^2$ as the space becomes flat \refs{\GepnerWI}.

It is important to note that even though we have just identified the 
flat space spectrum as arising from
the $w=0$ sector of the original theory, this does not mean that none of the
$w=0$ short strings remain in the plane wave.  
Some of the states can still carry 
$J \sim Q_5$, and as the limit $Q_5 \to \infty$ is taken we find the result
\efexp.  However, the $w=0$ plane wave states are generally farther 
from chiral than the spectral flowed states.

If $J^2/Q_5 \to 0$, then \efexp\ tells us that the string modes 
have energy that diverges as $\sqrt{Q_5}$.  
Note, however, that even in this case the 
supergravity modes (i.e. states at grade $1/2$ for both the right and 
left movers) remain, and they fall
into the global $\SL \times \SU$ multiplets.

\newsec{When the radius is small}

An extremely interesting question one would like to address is
what we can learn about string theory on $AdS \times S$ from
string theory on plane waves.  In the case of $\ads \times S^3$ and its
plane wave limit, we have a good understanding of both string theories,
and we now turn to this question.

But first, we'd like to stress a small point, which is that a priori 
there are two distinct
notions of ``high curvature'' one needs to keep in mind.  When one speaks
of a highly curved plane wave, that actually means 
\eqn\opplim{
\mu \apm p^+ >>1 \;.
}
In this case the string spectrum consists of highly spectral flowed states.
We see from \ejlim\ that this means the low lying string modes become
almost degenerate.  This is similar to what happens in the $AdS_5 \times S^5$
plane wave.  

Despite being ``highly curved'', the highly curved plane wave still involves
taking the radii of $AdS \times S$ to infinity.
Hence the GS superstring in highly curved plane waves is still amenable to 
quantization.
The second, and more interesting, notion of ``high curvature'' is obtained
by dropping the $R^2 \to \infty$ condition.  Then clearly the geometry cannot
be thought of as a plane wave.
Since it is only after the
Penrose limit is taken that the GS string can be solved, 
presently known 
results about the plane wave of $AdS \times S$ are not
expected to remain valid  in the case of small radius.     

However, there have been some reasons to think that the plane wave
spectrum \delj\ might continue 
to correctly
describe the large $J$ spectrum even outside the strict 
$Q_5 \to \infty$ limit.
Authors of \refs{\gms} studied various aspects of string theory on the
plane wave \ppmet\ from the point of view of the dual 
$({\cal M})^{Q_1Q_5}/S_{Q_1Q_5}$ CFT.  One of the more interesting things
they found in that work was that after extrapolating
\delj\ to $Q_5=1$, the result surprisingly agrees with the 
spectrum predicted by the dual
CFT at the orbifold point\foot{In fact, the NS $Q_5=1$ is the 
only case where a perfect agreement was 
found.  Matching of the spectra in general requires $g_s^2$ corrections 
and on the CFT side involves
moving away from the orbifold point in the moduli space.}.    
Since the CFT spectrum is believed to be reliable for arbitrary $Q_5$, 
whereas the string spectrum was found under the assumption that 
$Q_5$ is taken to infinity, this hints that 
perhaps \delj\ is true even when the spacetime geometry
does not correspond to a plane wave.   There have also been 
some work along this line
for the $AdS_5 \times S^5$ plane wave \refs{\BeisertTN, \BeisertTE}, but with
some differences, which we will discuss in the last section. 

We can answer this question directly for the $\ads \times S^3$ case
since we worked out the string spectrum that is
valid for all values of $Q_5$.  Our results apply equally to small $Q_5$,
when we should think of the geometry as $\ads \times S^3 \times {\cal M}$ 
with the first two factors being highly curved.
Thus, we can take \condl, \engiven\ and expanding for arbitrary fixed $Q_5$, 
large $w$, we find
that, in fact, the large $J$ spectrum is again given by \ejlim.  
We conclude that the plane wave spectrum is actually the large $J$ 
spectrum of strings on $\ads \times S^3 \times {\cal M}$, for arbitrary
values of the radius.

Actually, there are two special cases where the worldsheet description
we have given so far could break down.  These special cases occur for  
$Q_5=1$ or 2, whereby due to the shift in the level of the bosonic
WZW models the $\SU$ model acquires a negative or zero level.  However,
the problem is not serious for the $Q_5=2$ case as we can understand 
it to mean that only the fermionic
fields are present on the worldsheet for the $S^3$ part of 
the target space.    
The $Q_5=1$ case truly presents us with a difficulty since it is not
clear how to make sense of the $\SU$ WZW model at level $-1$ as a physical
theory.  It is not known at present how to describe the $Q_5=1$ model,
but arguments were presented in \refs{\SeibergXZ} to suggest that it is
a sensible (albeit very special) system.  
We'd like to argue that the result \ejlim\ is valid even for the $Q_5=1$ case
even though our starting point was not suited to describe it.
For one, it would be rather unusual for the expression \ejlim\ to be true for
$Q_5=2, 3, \dots \infty$ and not be true for $Q_5=1$ when nothing special
happens as we try to set $Q_5=1$.  More importantly, the dual CFT 
is well defined at $Q_5=1$ and its prediction for the string 
spectrum \refs{\gms} matches perfectly with \ejlim.  Perhaps $Q_5=1$ actually
represents the zero radius limit of $\ads$, thus providing the reason behind
perfect agreement with the symmetric orbifold.  The orbifold point of 
the CFT corresponds to the free theory (analogous to setting $g_{YM}=0$ 
in $AdS_5/{\rm CFT}_4$), whose dual string theory would apparently be
formulated on zero radius $\ads$.  We will return to this issue in the
Discussion.

\newsec{Summary and Discussion}

The two main objectives of this paper have been 
\item{(a)} To provide a CFT description of strings in a plane wave background,
giving the necessary framework for a detailed study of BMN 
correspondence using the powerful tools of CFT.
\item{(b)} To investigate the relationship between string theory
on $AdS \times S$ and string theory on plane waves, using the solvable
$\ads \times S^3$ case as a model.

\noindent
We offer some comments on each of these issues.

It is worth emphasizing that we are now positioned to take advantage of
the CFT techniques to study string interactions in the $\ads \times S^3$
plane wave.  This is in stark constrast
to the much studied case of the ten dimensional plane wave that arises
from $AdS_5 \times S^5$, where the RNS description of strings is lacking
and interactions can only be studied using string field theory.  
In fact, correlation functions in $\ads$ have already been 
calculated \refs{\mot},
so together with the correlation functions of $\SU$ WZW model
it should be possible to obtain scattering amplitudes in the plane
wave by appropriately taking the large $J$, $Q_5$ limit.  This should
prove to be an useful area for study.

In regards to the $\ads$ correlation functions, we show in Appendix A
that the spectral flow
number violation rule found in \mot\ can be understood as 
the conservation of angular momentum in the plane wave.

Additionally, one expects that
the map between the CFT operators and plane wave string states is easier 
to establish than the ten dimensional case,
owing to the fact that the $\ads/{\rm CFT}_2$ duality is highly constrained
by the infinite dimensional conformal symmetry.  
Thus, it becomes a technically simpler problem to study the BMN 
correspondence in situations where many interacting string modes 
are involved.      

The other main point of this paper is that we have actually compared
string theory on $\ads \times S^3$ to string theory on the plane wave.
We have found that the plane wave spectrum, which one might have
thought to be the result of some simplification of the $\ads \times S^3$
spectrum that occurs in the limit $Q_5, J \to \infty$, actually is
the result of $J \to \infty$ only.      

Recently it has been conjectured by Frolov and 
Tseytlin \refs{\FrolovQC, \FrolovTU, \TseytlinII} 
that the semi-classical formula for the energy of strings 
carrying spins in multiple directions in $AdS_5 \times S^5$ continues
to hold true at small values of the radius, provided that the spins take
very large values\foot{See \MateosDE\ for a discussion regarding the 
supersymmetry of the spinning strings.}.  Based on the findings of this 
paper, we feel strongly that their conjecture is true.  Furthermore,
if the relationship between string theory on $\ads \times S^3$ and its 
plane wave limit applies to other $AdS \times S$ spaces,
it suggests that the string spectrum on the plane wave limit of 
$AdS_5 \times S^5$ \refs{\MetsaevBJ ,\bmn} is in reality the large $J$
string spectrum on $AdS_5 \times S^5$.

Before leaving the subject of the Frolov-Tseytlin solution, let us note
a curious fact.  Frolov and Tseytlin found that 
the solution carrying two non-zero equal spins in
$S^5$ has the energy
\eqn\fte{
E=\sqrt{(2J)^2 +{R^4\over \alpha '^2}}\;.
}
This bears striking resemblance to the energy of a low-lying short string
state in $\ads \times S^3$ with the single spin
\eqn\resem{
E \sim \sqrt{J^2 +c {R^2 \over \alpha '}}\;,
}
where $c$ is a number of order 1.  Other than the difference in the power
of $R^2/\alpha '$, which could be explained by the fact that the role
of $N$ in $AdS_5/{\rm CFT}_4$ is played by both $Q_1Q_5$ and 
$\sqrt{Q_1Q_5}$ in $\ads/{\rm CFT}_2$ \refs{\gms},
the two expressions are almost identical.  It should be kept in mind that
\fte\ is a classical result whereas \resem\ is a quantum one.
It is not clear if Frolov-Tseytlin solution has an interpretation
as giving arise to a simpler spacetime geometry in a manner similar 
to BMN.  However, as we have seen, strings with large $J$ in 
$\ads \times S^3$ have a simple description even though it is only after the
radius is taken to be large as well 
that they can be viewed as moving in the plane wave.    
At any rate, it would be extremely interesting to 
understand why these two expressions
are so similar.  Perhaps studying strings on 
$\ads \times S^3 \times S^3 \times S^1$ \refs{\ElitzurMM}, which makes 
multi-spin solutions possible, along the lines of this
work will shed light on this issue\foot{The author would like to thank
A.~Adams for this point.}.

Another topic of interest has been pursued in 
\refs{\BeisertTN, \BeisertTE, \DharFI, \ClarkWK,\deMedeirosHR,\BonelliZU,
\SagnottiQA} 
involving
strings in the critical tension limit and the possibility of defining
string theory in the zero radius limit of $AdS$.  The hope is to
take the $\lambda \to 0, N \to \infty$ limit of $AdS/{\rm CFT}$ at 
its face value and
establish a duality between string theory in the zero radius $AdS$ and
a free field theory.  We should mention from the start, however,
that the approach has been to send $R^2/\alpha '$ to zero in the 
classical Hamiltonian and then quantize the resulting (simpler) theory.
This by no means assures us that we will find the same results when
we take the same limit in the quantum theory.  Another point to 
keep in mind is that when the radius of the spacetime is comparable to the
string scale, it is not clear whether one can even assign a definitive
value to the radius.  

Now we focus on the $\ads \times S^3$ example and try to address this issue.
Strictly speaking, one must set $Q_5=0$ to study the zero radius $\ads$.
In this case we do not know how to make sense of the worldsheet theory.
However, as stated above we do not believe that one should insist
on being able to set $R^2/\alpha '$ exactly to zero in the quantum treatment.
For the time being, we will be content with considering 
$R^2 \sim \alpha '$, which is still a nontrivial case. 
It is perhaps useful to recast the large $J$ expansion of the exact
energy formula found in section 5 using the radius of curvature in
string units (we ignore the internal space ${\cal M}$ for this discussion,
whose contribution is suppressed anyway):
\eqn\recast{
E - J =2+ q_{SL} + \bar{q}_{SL} + q_{SU} + \bar{q}_{SU}+
{R^2\over \alpha ' J}(N + \bar{N} + M + \bar{M} -2w -1) \;. 
}
We should note that the last terms in parantheses is what gives \recast\ its
stringy nature.  If for some reason (such as simply taking the 
``tensionless'' limit $R^2/\alpha ' = 0$ while continuing to trust \recast)
 they were absent, what remains would
resemble a field theory spectrum.  It might seem at first 
that the last terms would be negligible
for large $J$, finite $R^2/\alpha '$.  But in fact this is not the case,
because the excited string modes generically have grade
of order $\alpha 'J /R^2$ due to spectral flow.  
The only way in which the last terms in \recast\ disappear is in strict
${R^2/\alpha ' J} =0$ case\foot{Note that the combination $R^2/\alpha ' J$
is the square root of the coupling constant $\lambda '$ identified in
the BMN limit of $AdS_5 \times S^5$ \refs{\KristjansenBB, \ConstableHW}.}.  
When that happens the spectrum can be
schematically written 
\eqn\schem{
H_{lc} \sim \sum_{all \ modes}a^{\dagger}a \;,
}
which looks like a free field theory\foot{However, not all information about
string excitation numbers seems to be lost since the $L_0 =\bar{L}_0$ 
constraint still needs to be imposed.}.  
This suggests that the theory with
$R^2/\alpha ' =0$ (whatever its proper description might be) 
is not continuously connected to the $R^2 \sim \alpha'$ cases at finite $J$.

In a related topic, authors of {\BeisertTN, \BeisertTE} found evidence 
that the string
spectrum on the plane wave limit of $AdS_5 \times S^5$  may be extrapolated
down to finite $J$ after setting $g_s$ to zero, which has the effect of
reducing the spectrum to the form \schem.  The agreement with the SYM
prediction (which was done in \BeisertTE\ for conformal weights upto 10) 
as well as 
considerations of this paper lend support to the claim that in fact the
entire string spectrum on $AdS_5 \times S^5$ reduces to \schem\ at $g_s=0$.

There have also been some work on computing $R^{-2}$ corrections to
the plane wave spectrum as a way of approximating the $AdS \times S$ 
spectrum \refs{\CallanXR, \ParnachevKK, \ParnachevQS}.  
The results of this paper might be useful as a guide in checking
higher order calculations.  It is important to
note, however, that in computing corrections to the plane wave one does not
have the freedom to choose $R^2$ and $J$ independently.  The advantage
we had in the $\SL \times \SU$ model was being able to vary $Q_5$ and
$J$ in an independent manner.

In conclusion, strings in $\ads \times S^3$ and its plane wave or its 
large $J$ limit seem to be very useful models to study and it is hoped
that they will lead to a better understanding of the more complicated
plane wave/CFT and $AdS$/CFT dualities.

\

\centerline{\bf Acknowledgements}
The author thanks A.~Adams, J.~Maldacena, and S.~Minwalla for helpful
discussions.  This work was supported in part 
by DOE grant DE-FG01-91ER40654.

\appendix{A}{The spectral flow number violation rule}

In \refs{\mot} it was found that the $N$-point function of vertex 
operators with spectral flow numbers $w_i$, viewed as describing
the interaction of $i=2, \dots ,N$ 
incoming strings and $i=1$ outgoing string, vanishes unless\foot{The
discrete states are taken to be in the ground states of their
representations, 
i.e.~$\tilde{n}_i= \tilde{\ell}_i$.}
\eqn\specsat{
w_1 \le \sum_{i=2}^N w_i +N-2 \;.
}
This result was derived using representation theory of $\CSL$ algebra,
irrespective of what the spacetime consists of besides $\ads$,
and does not rely on any particular physical picture.  

We now show that when considering the plane wave limit of 
$\ads \times S^3 \times {\cal M}$, \specsat\ can be 
understood as enforcing the conservation of $J$.  In order
to find a non-zero correlation function the $J_i$ must
satisfy
\eqn\consj{
J_1=\sum_{i=2}^N J_i\;.
}  
We now divide both sides of this equation by $Q_5$ and identify
$w_i$ as the integer part of $J_i/Q_5$ (see the footnote below and
also note that we are in the $J, Q_5 \to \infty$ regime).  
On the RHS,
there will be  $N-1$ terms, each of the form $w_i + \Delta_i$
where $0 \le \Delta _i < 1$.  The sum of $\Delta_i$'s will therefore
be less than $N-1$.  Hence the spectral flow
numbers will satisfy \specsat.

\listrefs

\end